\documentclass[11pt]{article}
\usepackage[a4paper, margin=1in]{geometry}
\usepackage{graphicx}
\usepackage{amsmath, amssymb}
\usepackage{cite}
\usepackage{hyperref}
\usepackage{authblk}
\usepackage{caption}
\usepackage{subcaption}
\usepackage{booktabs}
\usepackage{float}
\usepackage{float}
\usepackage[utf8]{inputenc}
\usepackage{newunicodechar}
\newunicodechar{∼}{\textasciitilde{}}

\title{\textbf{ Classifying Radio-Loud and Radio-Quiet Quasars With Novel PCA Based Regression Classifier}}

\author{
Ramkrishna A. Joshi\textsuperscript{1} and  Vivek J. Shinde\textsuperscript{1}
}

\affil{\textsuperscript{1}\small Department of Physics, Indian Institute of Technology Hyderabad, Hyderabad, India}

\date{Dated: 30th April 2025}

\begin{document}

\maketitle

\begin{abstract}
The problem of quasar classification comes in the class of highly imbalanced classification problems since Radio-loud (RL) quasars are rare and make up only about 10\% of the quasar population. In this work,we use the Sloan Digital Sky Survey-DR3 dataset and introduce a  PCA-based regression pipeline designed to maximize recall for rare classes in class-imbalanced astronomical data. We demonstrate an effective methodology to identify the key features of the dataset and apply Principal Component Analysis (PCA) for dimensionality reduction. For the PCA transformed SDSS-DR3 dataset, first two components account for the 97\% of the observed variance. We perform classification of Radio-Loud (RL) and Radio-Quiet (RQ) quasars with Random Forest Classifier (RFC), novel PCA based balanced linear regression classifier (PBC), Random forest integrated with SMOTE classifier and XGBoost classifier with threshold tuning. RFC achieves an overall accuracy of 92\% while PBC achieves an overall accuracy of 62\% . XGBoost achieves an overall accuracy of 72\% and SMOTE integrated RFC achieves an accuracy of 85\%. Higher precision is obtained for RQ quasars in all classification methods. For the RL class, RFC achieves a recall of 0.04, XGBoost achieves a recall of 0.39, SMOTE integrated RFC achieves a recall of 0.25 and PBC achieves a recall of 0.52 attributed to the balanced logistic regression. RFC and PBC achieve F1 score of 0.08 and 0.19 respectively for RL while XGBoost achieves an improved F1 score of 0.22 but at the cost of reduced recall of the RL class. SMOTE integrated RFC achieves a better F1 score of 0.21 over RFC and PBC .Overall results of classifiers point to extreme class imbalance between RQ and RL classes in the data set. This PCA-driven approach, in combination with machine learning and statistical testing, provides a robust framework for distinguishing radio-loud from radio-quiet quasars based on their optical photometric and redshift properties. Additionally, we also investigate the optical photometric properties of radio-loud and radio-quiet quasars using the u, g, r, i, and z band magnitudes and demonstrate the Lyman-Alpha forest effect for high z quasars with $z>4$. We show a significant difference in the redshift distributions of RL and RQ.
\end{abstract}


\section{Introduction}
Quasars or quasi-stellar objects, are among the brightest and energetic objects in the universe. They are distant objects powered by the accretion of material onto supermassive black holes at the centers of galaxies, releasing immense energy across the electromagnetic spectrum \cite{ref1}. Studying quasars provides critical insights into the early universe, black hole growth, and the co-evolution of galaxies \cite{ref2}. With the advent of large sky surveys, such as the Sloan Digital Sky Survey (SDSS), the astronomical community now has access to vast catalogs of quasar observations \cite{ref3}.The Sloan Digital Sky Survey (SDSS) has cataloged thousands of quasars, providing a large scale photometric and x ray data. Specifically, it has enabled detailed studies of quasar distributions, spectral properties, and redshift distributions. One of the major challenges in quasar research is distinguishing between radio-loud and radio quiet quasars. Radio loud quasars emit strong radio waves, likely due to relativistic jets from the active galactic nucleus (AGN), while radio-quiet quasars have weak or no such emissions. Radio quiet quasars constitute the majority of the quasar population, while radio loud quasars are relatively rare. This imbalance is often attributed to the presence or absence of powerful relativistic jets, which are the primary source of strong radio emissions. The formation of such jets is believed to depend on several factors, including the spin of the central supermassive black hole (SMBH) \cite{ref21}. Higher black hole spin can lead to more efficient jet production, thus increasing the likelihood of a quasar being radio-loud \cite{ref22}. Additionally, the environment surrounding the black hole, such as the density of the interstellar medium, can influence jet formation and propagation. Consequently, only a subset of quasars develop the necessary conditions to become radio-loud, making them an interesting subject for understanding the mechanisms behind jet formation and AGN feedback \cite{ref20}. Given this intrinsic imbalance, classifying quasars based on their radio properties presents both a scientific and computational challenge. Optical and radio features, while individually informative, often exhibit significant overlap and correlation, making it difficult to distinguish between the two classes using raw observational data alone. This motivates the use of statistical and machine learning techniques to reveal patterns that may not be immediately evident. In this study, we use data from the SDSS quasar catalog to explore whether photometric features can be effectively used to classify quasars as radio-loud or radio-quiet. To enhance model performance and reduce dimensionality, Principal Component Analysis (PCA) is first applied to compress the optical magnitude data and training regression-based classifier, followed by the use of a Random Forest Classifier and XGBoost to build a robust and interpretable classification model. Previous works have focused on classification of quasars using KNN, Gradient boosting algorithms, Random forests, NNs and MLPs \cite{ref23} \cite{ref24}. Significant amount of work is done in applying PCA to quasars for identifying principal components associated with physics properties of QSOs \cite{25}. In this work we present a novel PCA based balanced linear regression classifier and compare its performance to the threshold tuned XGBoost and Random forest classifier.

Principal Component Analysis (PCA) is a widely used technique for dimensionality reduction, which transforms a set of possibly correlated variables into a set of linearly uncorrelated components known as principal components . In physical terms, this can be treated equivalent to determining the principal axes of the system. This technique improves interpretability of the data while minimizing the information loss \cite{ref4}. A crucial feature is standardization of the data to ensure that each feature contributes equally to the analysis. For a feature \( x_j \), this is achieved by
\[
x_j^{\text{(std)}} = \frac{x_j - \bar{x}_j}{\sigma_j},
\]
where \( \bar{x}_j \) is the mean and \( \sigma_j \) is the standard deviation of feature \( j \). After standardization, the covariance matrix \( \Sigma \in \mathbb{R}^{p \times p} \) of the dataset is computed:
\[
\Sigma = \frac{1}{n-1} X^\top X,
\]
where \( X \in \mathbb{R}^{n \times p} \) is the standardized data matrix. PCA then performs an eigen decomposition of \( \Sigma \), giving eigenvalues \( \lambda_1, \lambda_2, \ldots, \lambda_p \) and corresponding eigenvectors \( \mathbf{v}_1, \mathbf{v}_2, \ldots, \mathbf{v}_p \). Each eigenvector defines the direction of a principal component, and its associated eigenvalue quantifies the variance explained by that component.

The proportion of total variance explained by the \( i \)-th principal component is given by the explained variance ratio:
\[
\text{Explained Variance Ratio}_i = \frac{\lambda_i}{\sum_{j=1}^{p} \lambda_j}.
\]
To determine the number of components needed to retain a specific amount of variance (e.g., 97\%), we compute the cumulative explained variance:
\[
\text{Cumulative Variance}_k = \sum_{i=1}^{k} \frac{\lambda_i}{\sum_{j=1}^{p} \lambda_j}.
\]
For example. the smallest number \( k \) satisfying
\[
\sum_{i=1}^{k} \frac{\lambda_i}{\sum_{j=1}^{p} \lambda_j} \geq 0.97
\]
indicates the minimum number of components required to preserve 97\% of the dataset’s variance \footnote{Refer \href{https://www.microsoft.com/en-us/research/wp-content/uploads/2006/01/Bishop-Pattern-Recognition-and-Machine-Learning-2006.pdf}{"Pattern Recognition and Machine Learning"} by Christopher M. Bishop, Chapter 12, p 563-565 for a more formal derivation as above. For a more mathematically rigorous derivation refer \href{https://www.deeplearningbook.org/contents/linear_algebra.html}{"Deep Learning"} by  Ian Goodfellow, Yoshua Bengio, and Aaron Courville, Chapter 2, p 45-50.}.
\vspace{0.3cm}

The practical application of the dimensionally reduced data is to facilitate the task of radio-loud and radio-quiet quasar classification with the aid of classifier frameworks from machine learning.

Random Forest is an ensemble learning method that constructs multiple decision trees during training and outputs the class that is the mode of the classes (classification) of the individual trees. Random forests form an ensemble of tree predictors with each tree attributed to random vectors sampled independently with same underlying distribution for all trees \cite{ref5}.It reduces overfitting by averaging multiple deep decision trees trained on different parts of the same dataset . Each tree is built from a bootstrap sample and a random subset of features is considered for splitting at each node \cite{ref6}. This makes the classifier robust to noise and effective for high-dimensional data.

Another popular algorithm is XGBoost (Extreme Gradient Boosting).XGBoost is an efficient and scalable version of Gradient Boosting (GB) algorithms. It has gained popularity due to its performance in machine learning tasks, particularly in structured/tabular data. Gradient boosting algorithms work on the principal of iterative refinement of the previous model. This is somewhat similar in context to how with Euler algorithm, we predict the outcome at the next time step by using the previous outcome. Similarly XGBoost initially trains a shallow decision tree for predicting the outcome. This initial decision tree is susceptible to errors in the prediction. The next step involves building a decision tree with the previous error information fed into it. The next decision tree is able to improve upon the prediction by using the previous tree errors and predicting the outcome closer to the true outcome. This process is iteratively carried out until the prediction matches the expected outcome. Each new tree corrects the mistakes of the previous and takes the model one step closer to the true distribution. XGBoost implements a boosting algorithm where weak learners (usually decision trees) are combined to create a stronger predictive model with an additional L1(Lasso) and L2(Ridge) regularization for handling tree weights effectively\cite{ref19}. This method effectively avoids the risk of overfitting. Another feature of XGBoost that allows for a faster computation is the parallel decision building. Conventional GB algorithms build each decision tree sequentially but XGBoost builds them in a parallelized fashion thus allowing for more computational output in lesser time. Additionally, XGBoost is a sparsity aware algorithm and can learn to handle missing values without any manual instructions. \footnote{XGBoost is a python package. Documentation of XGBoost usage can be found at \href{https://xgboost.readthedocs.io/en/stable/python/python_intro.html}{XGBoost documentation}.}

\vspace{0.3cm}

Another classification class implemented in the paper is PCA based balanced regression classifier. In this classifier model, PCA is first applied to reduce the dimensionality and noise in the data. The transformed data is then fed into a regression model, with class balancing techniques (such as sample weighting or resampling) to address imbalanced datasets. Class balancing is highly necessary for this dataset due to inherent class imbalance between radio loud and radio quiet quasars. Without class balancing, the classifier performs poorly on identifying the minority class, as demonstrated in the later sections of this paper. This approach improves both interpretability and performance \cite{ref7},\cite{ref8},\cite{ref9}.

\section{Methodology}

We utilize SDSS-DR3 dataset for the analysis.This work applies a structured data analysis framework to identify the key features of the SDSS catalog and develop a classification workflow to classify quasars into radio-loud and radio-quiet categories \cite{ref26}. The dataset includes photometric measurements in five optical bands (u, g, r, i, z) along with radio flux values. To better understand the diversity within the quasar population, we first explored the dataset’s basic characteristics:

\begin{itemize}
    \item A redshift distribution plot is used to understand the underlying density distribution and  highlighting high-redshift quasars (e.g., z $>$ 4).
    \item A correlation matrix across optical bands (u, g, r, i, z) helps visualize relationships between photometric bands, This guides in application of PCA to the data for dimensionality reduction. 
    \item A color-coded scatter plot helps identifying high-redshift and radio-loud quasars from the catalog.
\end{itemize}

These initial visualizations play a critical role in determining the key contribution features of the data to facilitate PCA and binary classification models In this study, we aim to classify quasars as radio-loud or radio-quiet using their photometric and radio properties. This work focuses on three key areas:
\begin{itemize}
    \item An in depth evaluation of the dataset to identify key features.
    \item Principal component analysis (PCA) for dimensionality reduction.
    \item Classification models to distinguish radio loud quasars from radio quiet quasars including a PCA based logistic regression classifier.
\end{itemize}

\begin{figure}[H]
    \centering
    \includegraphics[width=0.5\linewidth]{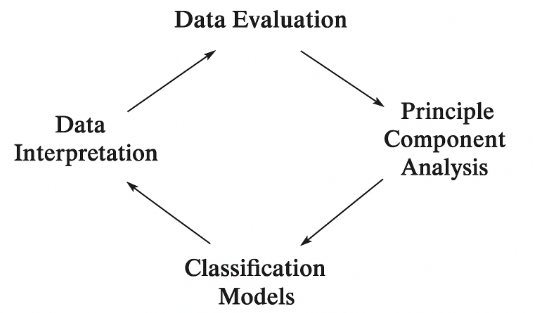}
    \caption{Methodology adapted for the SDSS data analysis.}
    \label{fig:enter-label}
\end{figure}

The key component analysis involves visualizing the redshift distribution to understand the range and density of observed quasars. This is implemented using the \texttt{seaborn.histplot()} function with the \texttt{kde=True} argument in Python, which automatically computes and plots the KDE curve. High-redshift sources ($z > 4$) were identified through scatter plots, and a correlation matrix of photometric bands was used to assess feature relationships and guide dimensionality reduction. Color-color pairplot method can be implemented in python by \texttt{seaborn.pairplot()} package. We also identify outliers from the data set with $i>19$ by simply by separating the outliers from $i<19$ dataset. We also perform Hypothesis testing to compare the redshift distributions of radio-loud and radio-quiet quasars. Using \texttt{scipy.stats}, a two-sample independent t-test (\texttt{ttest\_ind}) was applied. The tests evaluate whether the two samples come from populations with the same mean or distribution, respectively. A significance threshold of 0.05 was used to interpret the p-values and assess statistical differences between the two quasar populations. Pre-processing steps included handling missing values and z-score normalization to generate a standardized data. To define the classification target, we applied a logarithmic transformation to the radio flux and labeled sources with $log10(radio \ flux) > 1$ as radio-loud and the rest as radio-quiet \cite{ref10}. These labels formed the basis of our binary classification models such as RFC and PBC.\\

To highlight key variance features in the data, Principal Component Analysis (PCA) was applied to the standardized optical magnitudes. First, the photometric features were standardized using the \texttt{StandardScaler} from \texttt{scikit-learn} to ensure zero mean and unit variance. This step is crucial for PCA, which is sensitive to the scale of input  \cite{ref11}. In practice, PCA analysis can be implemented in Python using the \texttt{scikit-learn} library. The \texttt{PCA()} class provides the attribute \texttt{explained\_variance\_ratio\_}, which stores the explained variance ratios \( \left[ \frac{\lambda_1}{\sum \lambda}, \frac{\lambda_2}{\sum \lambda}, \ldots \right] \). The cumulative explained variance is computed as:
\begin{verbatim}

from sklearn.decomposition import PCA

pca = PCA()
pca.fit(X_std)  ## X_std is the standardized data matrix
cumulative_variance = numpy.cumsum(pca.explained_variance_ratio_)
n_components_K = numpy.argmax(cumulative_variance >= K) + 1 
## K is required explained variance 
\end{verbatim}

This approach allows for the dimensionality reduction of the data while preserving its structure and the variability. Additionally we study the PCA component loadings for each photometric band to interpret the contribution of each band. Finally, these PCA components were used to train a balanced logistic regression classifier.\\

The Random Forest classifier was implemented using \texttt{RandomForestClassifier} from the \texttt{sklearn.ensemble} module. The photometric features ($u$, $g$, $r$, $i$, $z$) were used as input variables, and the target was the binary classification of radio-loudness. The data was split into training and testing sets using \texttt{train\_test\_split}. A forest of 100 decision trees was trained, and the model's performance was evaluated using accuracy and a classification report on the test set. \\

We implement XGBoost in Python, by using \texttt{xgboost} package with scale factor set at the imbalance ratio. Let $N_0$ and $N_1$ represent the number of samples in class 0 (majority) and class 1 (minority), respectively. The imbalance ratio $r$ is then given by $r = \frac{N_0}{N_1}$. In Python, this can be computed using the \texttt{collections.Counter} class. This ratio can be used to set the \texttt{scale\_pos\_weight} parameter in XGBoost to give higher importance to the minority class during training. This approach provides a scalable analysis framework for interpretation of the astrophysics data under consideration.\footnote{Documentation of all the scikit methods and packages used is available online at \href{https://scikit-learn.org/stable/}{SciKit Documentation}}

\section{Results}

\subsection{Redshift distribution and kernel density estimation}

The distribution of redshift ($z$) for quasars in the dataset is visualized using a histogram with an overlaid Kernel Density Estimate (KDE). Comparison of KDE with gaussian and exponential likelihood provides insights into the effectiveness of each chosen likelihood. The KDE provides a smoothed approximation of the underlying probability density function of the redshift values, offering insight into the continuous structure of the distribution. The histogram shows the frequency of quasars within specified redshift bins, while the KDE highlights trends and peaks in the data without relying on bin width. Together, these provide a comprehensive view of how redshift values are distributed among the observed quasars.

\begin{figure}[H]
    \centering
    \begin{minipage}{0.5\linewidth}
        \centering
        \includegraphics[width=\linewidth]{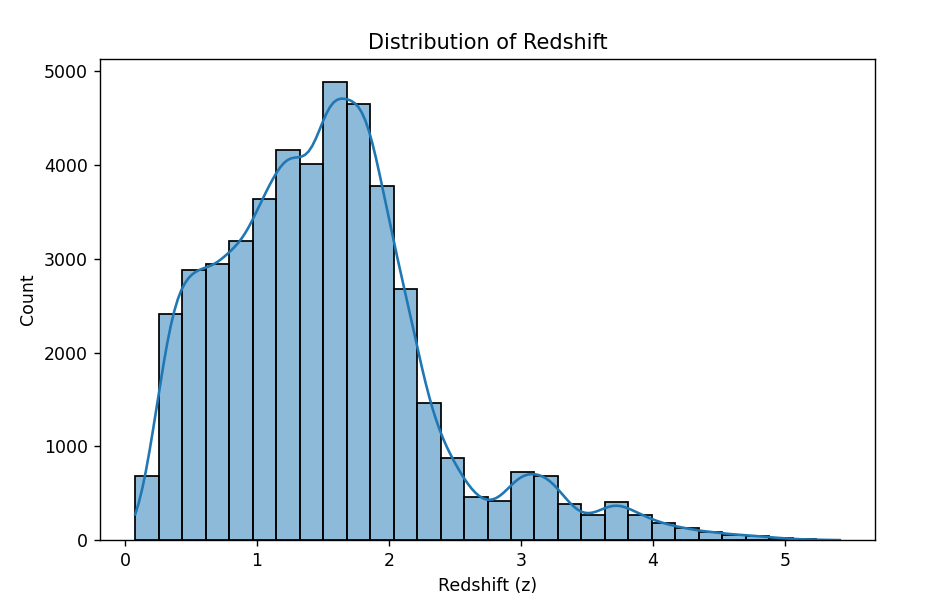}
        \textbf{(a)}
    \end{minipage}
    
    \vspace{0.5cm}
    
    \begin{minipage}{0.5\linewidth}
        \centering
        \includegraphics[width=\linewidth]{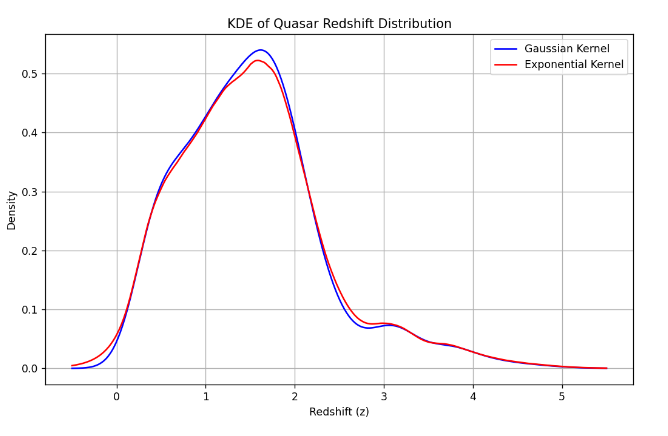}
        \textbf{(b)}
    \end{minipage}
    
    \caption{ Kernel Density Estimation with bandwidth of 0.2 for the Redshift of quasars from the SDSS data. (a) shows the histogram redshift KDE and (b) shows the estimation of KDE Guassian kernel (blue) and exponential kernel (red).}
    \label{fig:kde-comparison}
\end{figure}

The hypothesis testing result with the T-test P value of 0.00003 helps in rejecting the null hypothesis $H_{0}:$There is no significant difference in the redshift distributions of the radio-loud and radio-quiet quasars.

\subsection{Correlations Matrix for photometric bands}

To check for the credibility of performing PCA on the dataset, valuable information is gained through the photometric band magnitudes. Correlations between different bands are key features aiding this analysis. There are five major photometric bands under consideration namely u,g,r,i and z. We perform correlation analysis for each permutation and combination with these five bands. We observe strong correlation trends across various combinations.

\begin{figure}[H]
    \centering
    \includegraphics[width=1.1\linewidth]{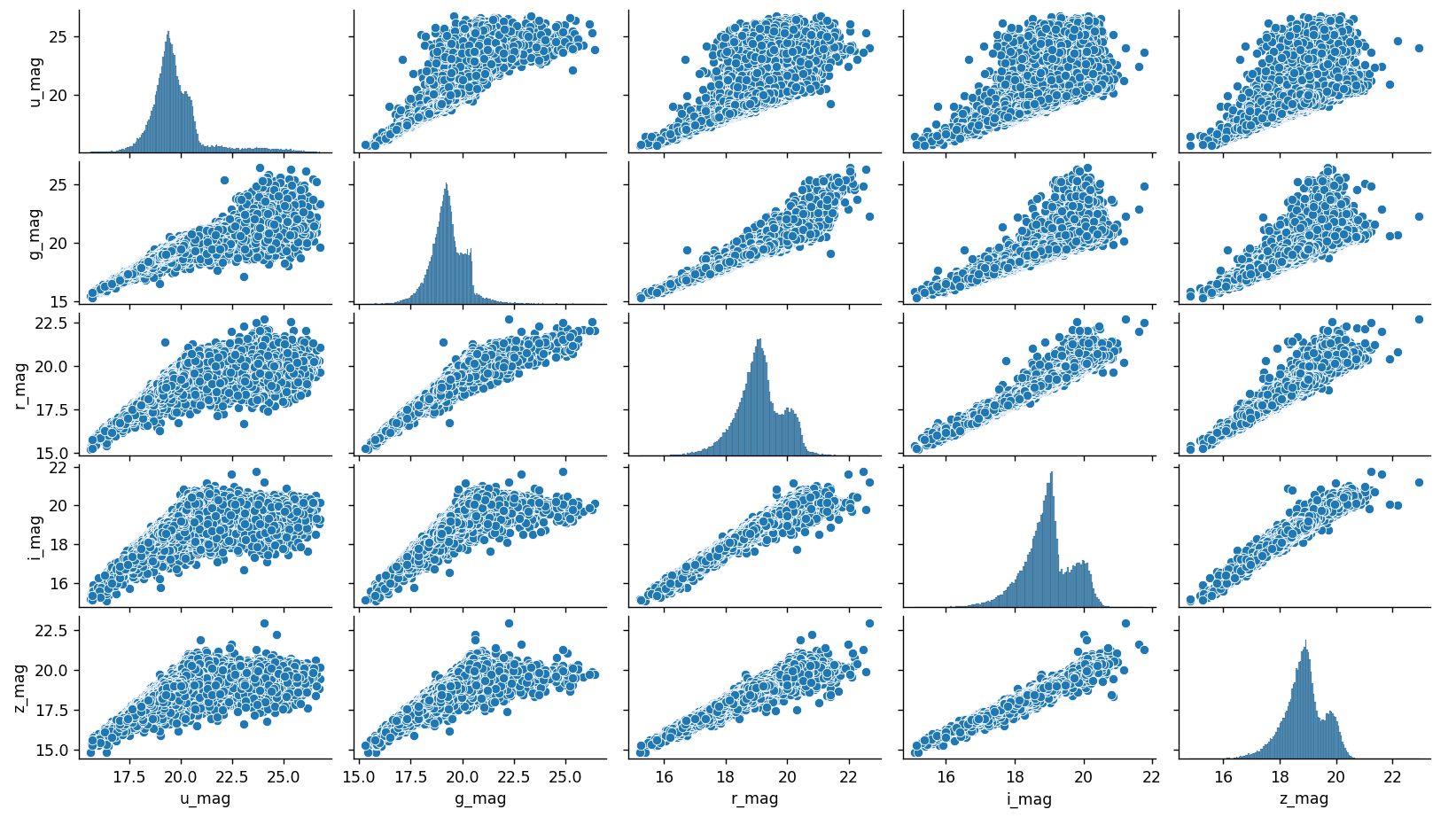}
    \caption{Figure shows pair-plot matrix for photometric band combinations. Y and X axis represent u,g,r,i and z magnitudes from top to bottom and left to right respectively. Diagonal elements represent distribution within each band while off-diagonal elements represent inter band relationships.}
    \label{fig:enter-label}
\end{figure}

There is a strong positive correlation for many band combinations however there are outliers in some of the bands which cause a significant deviation in the scatter plots. The pair plot visualizes the relationships among the SDSS photometric bands (u, g, r, i, z). Each off-diagonal plot shows a scatterplot between two bands, revealing strong linear correlations. The diagonal plots show the distribution of magnitudes in each band, which are roughly Gaussian but show some skewness. Outliers are visible in the scatter plots, particularly for sources with $i_{mag} > 19$, which may indicate either high-redshift quasars or noisy measurements. The number of outliers with $i_{mag} > 19$ are found to be 21685.

\subsection{Lyman Alpha forest effect and high z quasars}

High-redshift quasars, particularly those at redshifts $z\approx6$, provide crucial insights into the state of the intergalactic medium (IGM) and the process of reionization. These quasars, residing in dense regions of the IGM, exhibit increasingly thick Lyman-alpha ($Ly\alpha$) forests in their absorption spectra, which suggest that the fraction of neutral hydrogen in the IGM increases as the universe evolves towards higher redshifts \cite{ref12},\cite{ref13}. Their observation also provides crucial insights into the early universe. They provide valuable information about the growth of supermassive black holes in the universe and their relation to the formation of early galaxies. \cite{ref2}. These quasars are one of the most effective tools to probe the reionization epoch. In this work,we isolate and mark these high z quasars in a color-color plot.

\begin{figure}[H]
    \centering
    \includegraphics[width=1\linewidth]{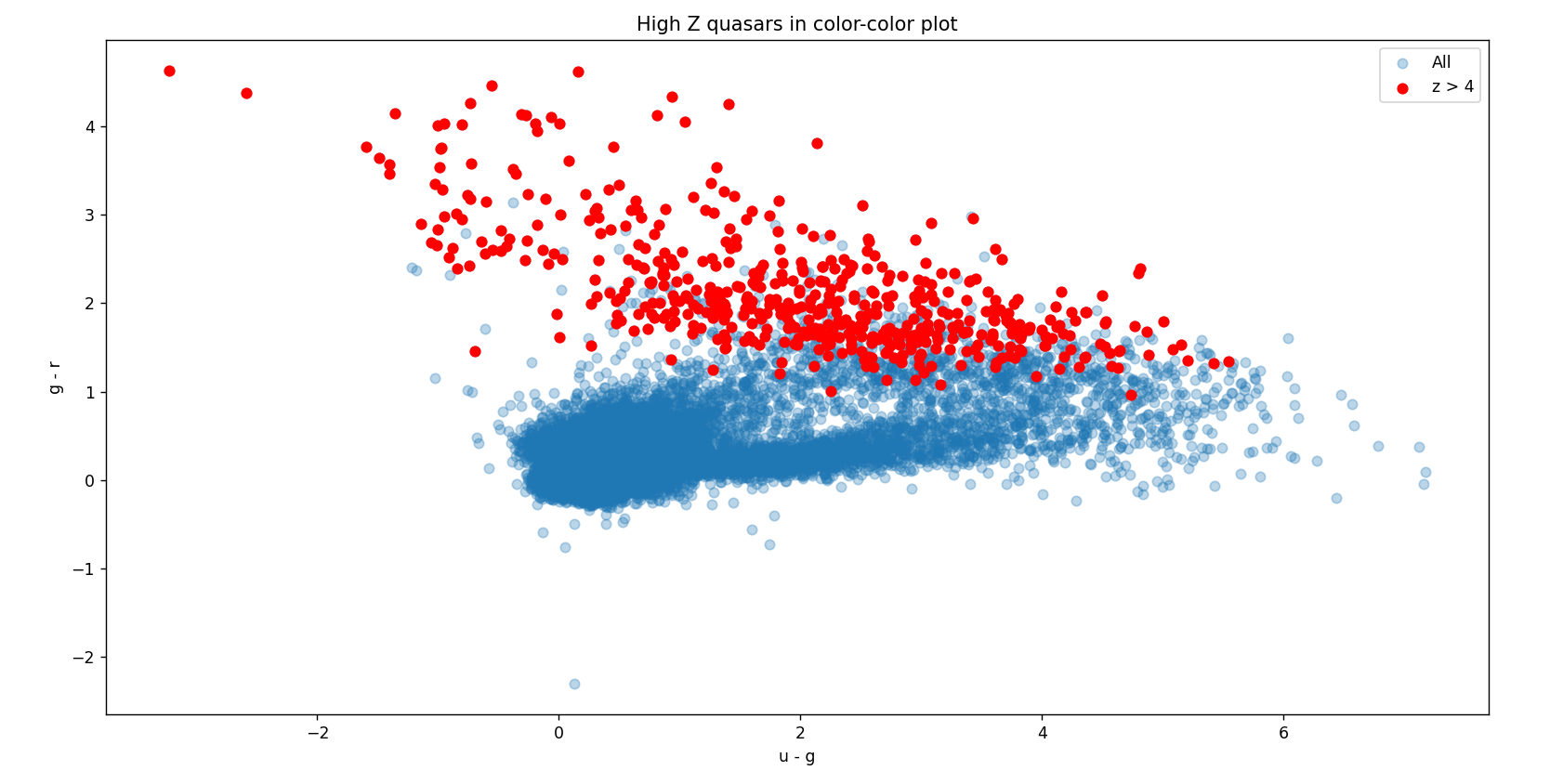}
    \caption{Figure shows the classification of $z>4$ quasars, marked in red, in the ug-gr color color diagram.}
    \label{fig:enter-label}
\end{figure}

High z quasars occupy a distinct region in the color-color plot above. Their g-r magnitudes are typically higher than 1. This provides a simple classification criterion of identification \cite{ref14}.

\begin{figure}[H]
    \centering
    \includegraphics[width=1\linewidth]{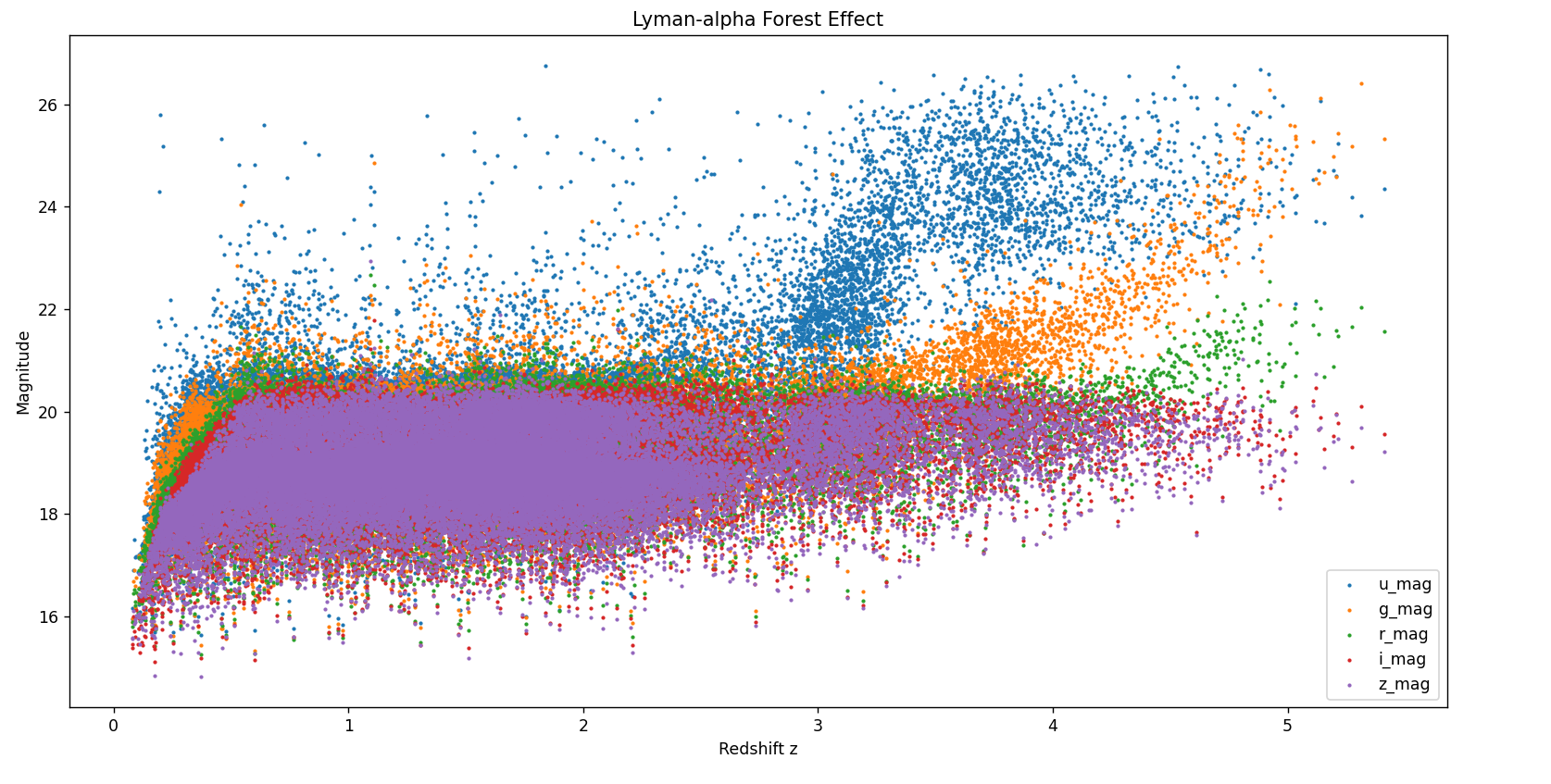}
    \caption{Figure shows the variation of SDSS photometric magnitudes ($u$, $g$, $r$, $i$, $z$) with redshift $z$ for quasars. In the figure above, the noticeable increase in $u$ and $g$-band magnitudes beyond $z \sim 2.5$ illustrates the impact of the Lyman-alpha forest absorption}
    \label{fig:enter-label}
\end{figure}

As redshift increases, intervening hydrogen clouds in the intergalactic medium absorb more ultraviolet light, leading to a suppression of flux in the bluer bands. This effect is particularly strong in the $u$-band, making it a useful feature for identifying high-redshift quasars \cite{ref15}. The Lyman-alpha forest is characterized by neutral hydrogen $(H I) \ Ly \alpha$ 1215.67 Å transitions and is sensitive to the temperature and density of the intervening gas \cite{ref12}.

\subsection{Identification of radio loud quasars}

\begin{figure}[H]
    \centering

    % Top row: two images side by side
    \begin{minipage}{0.45\linewidth}
        \centering
        \includegraphics[width=\linewidth]{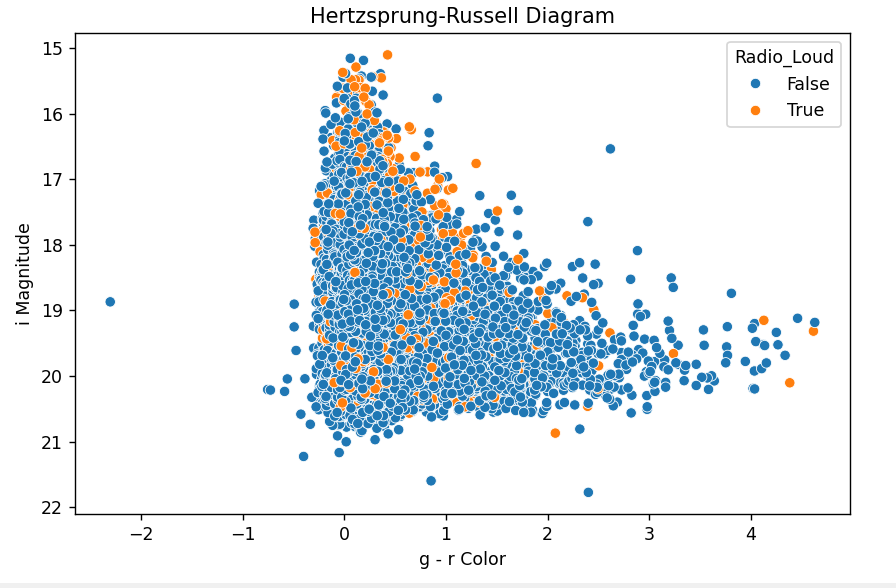}
        \textbf{(a)}
        \label{fig:gr_color_imag}
    \end{minipage}
    \hfill
    \begin{minipage}{0.45\linewidth}
        \centering
        \includegraphics[width=\linewidth]{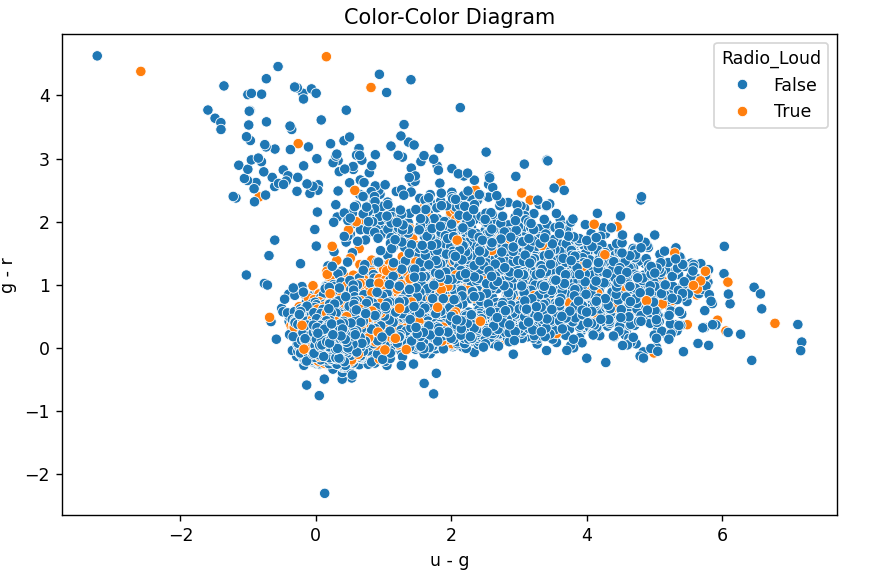}
        \textbf{(b)}
        \label{fig:gr_vs_ug}
    \end{minipage}

    \vspace{0.5cm}

    % Bottom row: single centered image
    \begin{minipage}{0.5\linewidth}
        \centering
        \includegraphics[width=\linewidth]{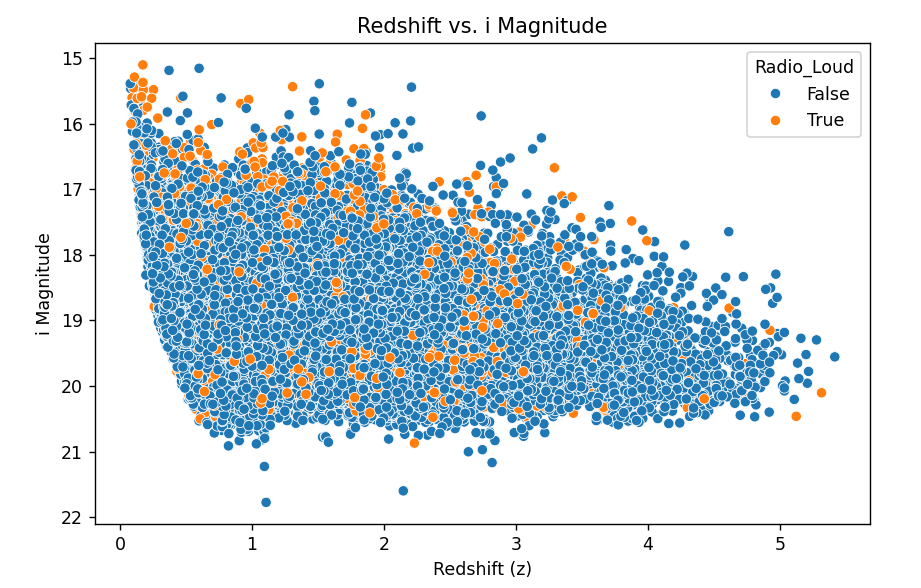}
        \textbf{(c)}
        \label{fig:imag_vs_z}
    \end{minipage}

    \caption{(a) $g - r$ color vs. $i$-band magnitude. (b) $g - r$ vs. $u - g$ color-color diagram. (c) $i$-band magnitude as a function of redshift. These plots help distinguish radio loud quasars from radio quiet ones.}
    \label{fig:color_selection_combined}
\end{figure}

Plots from the figure above are useful for photometric identification of quasars. In the previous section we have demonstrated the identification of high z quasars in the ug-gr color color diagram. High-redshift quasars tend to occupy a distinct region in these diagrams due to their redder colors caused by Lyman-alpha forest absorption and intrinsic spectral properties. For training PBC it is necessary to study features which can be useful for distinguishing radio-loud quasars from radio-quiet ones with optimal computational resources. Diagnostic plots as sbown above help in feature selection. Unlike high-z quasars occupying a distinct region in the color-color diagram, there is an isotropic overlap of loud and quiet quasar populations in all the figures above and hence a more robust criteria is required for feature selection.

\subsection{RFC and PBC analysis}

The SDSS catalog includes significantly large number of features explaining the variance in the data. Without the PCA, analysis of the data is cumbersome and highly computationally intensive task.

\begin{figure}[H]
    \centering
    \includegraphics[width=1.1\linewidth]{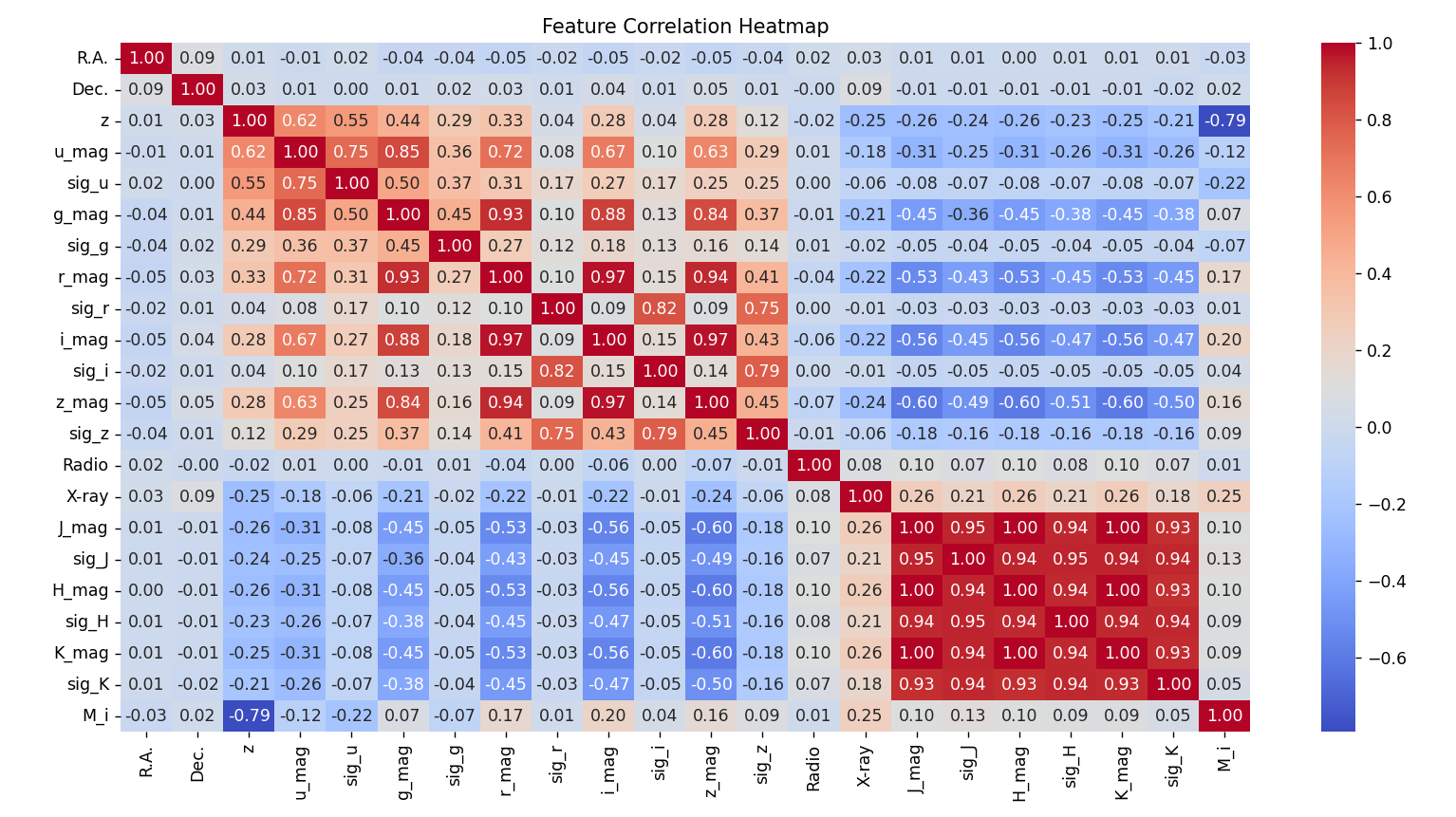}
    \caption{Figure shows feature correlation heatmap prior to PCA for all available features in the SDSS quasar catalog.}
    \label{fig:enter-label}
\end{figure}

The  heat map is a  $22 \times 22$ grid of Pearson correlation coefficients among all available features in the SDSS quasar catalog before dimensionality reduction. There are number of nuisance parameters which don't really play a role in the present analysis but nonetheless are important to be considered for quasars. Hence it is necessary to find an effective grid of parameters that essential for the analysis. This task is accomplished by PCA. Optical magnitudes across the u, g, r, i, and z bands show strong positive correlations, which suggests significant redundancy in the photometric data. Furthermore, the measurement uncertainties ($sig_u$, $sig_g$, etc.) are showing internal correlations, further contributing to the feature overlap. Notably, radio and X-ray measurements exhibit weaker correlations with the optical bands, which highlights their potential to act as a separate dimension in the PCA. The high degree of correlation among optical parameters may make it challenging for the classification models to decide on weight factors to be assigned to each parameter and hence render them ineffective \cite{ref16}. Therefore, dimensionality reduction becomes a necessary step for simplifying the dataset while preserving its essential structure. Therefore, applying PCA becomes essential to create a more stable, efficient,easy to handle and interpretable classification framework for distinguishing between radio-loud and radio-quiet quasars.\\

PCA is applied to the SDSS photometric magnitudes ($u$, $g$, $r$, $i$, $z$) of quasars. For the dataset, $k_{min}$ is found to be 2 to explain 97\% of the variance and 5 to explain $\approx 100 \%$ of the variance. The component loadings, presented in Table~\ref{tab:pca-loadings}, show the relative contribution of each magnitude band to the principal components.

\begin{table}[H]
\centering
\caption{PCA component loadings for SDSS photometric magnitudes}
\label{tab:pca-loadings}
\begin{tabular}{lccccc}
\toprule
\textbf{} & \textbf{u\_mag} & \textbf{g\_mag} & \textbf{r\_mag} & \textbf{i\_mag} & \textbf{z\_mag} \\
\midrule
\textbf{PCA1} & 0.391 & 0.459 & 0.468 & 0.462 & 0.451 \\
\textbf{PCA2} & 0.802 & 0.233 & -0.195 & -0.323 & -0.400 \\
\textbf{PCA3} & -0.428 & 0.686 & 0.311 & -0.181 & -0.465 \\
\textbf{PCA4} & -0.122 & 0.450 & -0.510 & -0.410 & 0.596 \\
\textbf{PCA5} & -0.071 & 0.247 & -0.622 & 0.694 & -0.256 \\
\bottomrule
\end{tabular}
\vspace{1cm}
\centering
\caption{PCA component variances and cumulative variances}
\label{tab:pca-variances}
\begin{tabular}{lcc}
\toprule
\textbf{Principal Component} & \textbf{Variance (\%)} & \textbf{Cumulative Variance (\%)} \\
\midrule
\textbf{PCA1} & 0.8759 & 0.8759 \\
\textbf{PCA2} & 0.0969 & 0.9728 \\
\textbf{PCA3} & 0.0183 & 0.9911 \\
\textbf{PCA4} & 0.0061 & 0.9972 \\
\textbf{PCA5} & 0.0028 & 1.0000 \\
\bottomrule
\end{tabular}
\end{table}

The first principal component (PCA1) exhibits uniformly positive loadings across all five bands indicating that PCA1 primarily captures the overall brightness of the sources. This component account for nearly 88\% of the variance in the data thus representing a global trend in the photometric data. The second component (PCA2) is dominated by a strong positive loading in the $u$-band (0.802), and moderate to negative contributions from the other bands. This suggests that PCA2 is sensitive to variations in ultraviolet emission, potentially reflecting differences in spectral slope or UV excess commonly associated with quasars \cite{ref17}, \cite{ref18}. 

\begin{figure}[H]
    \centering
    \includegraphics[width=0.9\linewidth]{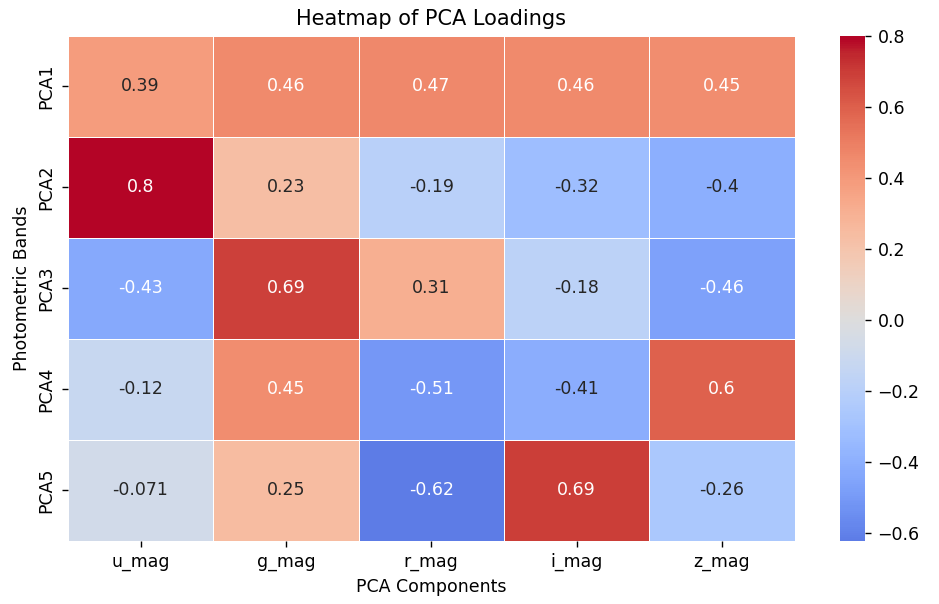}
    \caption{Heat map representation of contribution of each photometric band to PCA components. PCA 1 component has uniform loadings from all photometric bands. PCA 1 and PCA 2 components account for 97\% of the variance.}
    \label{fig:enter-label}
\end{figure}

PCA3 displays a significant positive loading in the $g$-band (0.686), with small negative contributions from the $u$ , $i$ and $z$ bands, accounting for model sensitivity to color variations. PCA4 and PCA5 capture more subtle variations in the dataset. The first two PCA features represent the majority of the explained variance. PCA has effective application for this dataset. The component explained variance is the highest for PCA 1, indicating that most of the data variation is captured by PCA 1. Hence a uniform superposition of all bands output, i.e the overall brightness is the key feature of the dataset. Other PCA components explain minor variations in the data accounting for complexity of the dataset. For the majority of the data variance to be captured by this method, one would only requires two components. This high dimensional data can be effectively modeled by just a 2D data through application of PCA. 

\begin{figure}[H]
    \centering
    \includegraphics[width=0.9\linewidth]{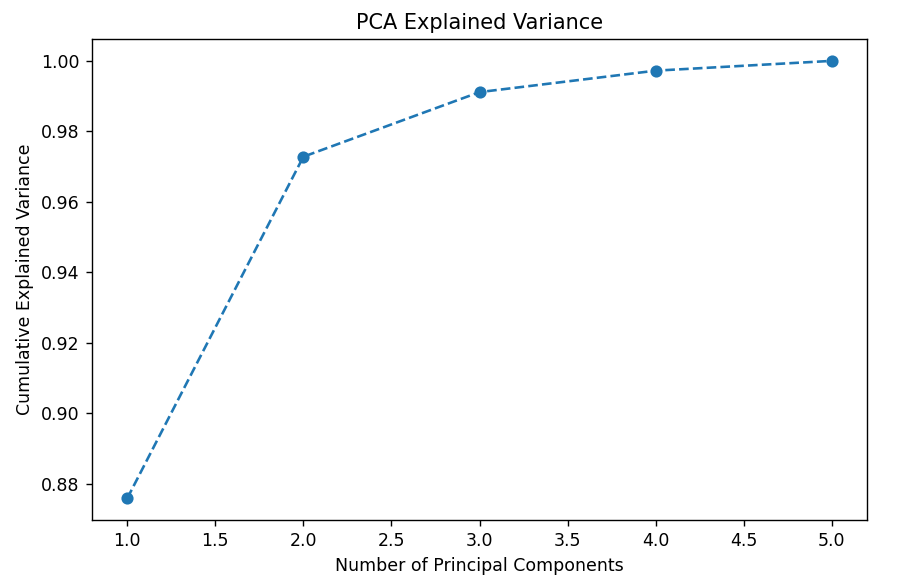}
    \caption{Figure shows the variation of cumulative explained variance against the number of principal components considered. PCA 1 explains about 88\% of the variance while the first two components (PCA 1 and PCA 2) explain 97\% of the variance. Five PCA components are required to explain the entire data.}
    \label{fig:enter-label}
\end{figure}

As observed in the data output from Table~\ref{tab:pca-variances}, the cumulative explained variance curve for the principal component analysis (PCA), shown in Figure above, illustrates the proportion of total cumulative explained variance as a function of the number of principal components included. The first component alone accounts for approximately 88\% of the total variance as claimed. Majority of the information present in the photometric magnitudes can be captured by a single linear combination of the original features. The addition of the second component raises the cumulative explained variance to over 97\%. The curve asymptotically approaches unity, with the fourth and fifth components contributing marginally to the total variance. This behavior confirms that the data possess a strong low-dimensional structure. PCA loadings indicate to the fact that the quasar photometric data is dominated by gross brightness and a small number of color-dependent variations.

\begin{figure}[H]
    \centering
    \begin{minipage}{0.7\linewidth}
        \centering
        \includegraphics[width=\linewidth]{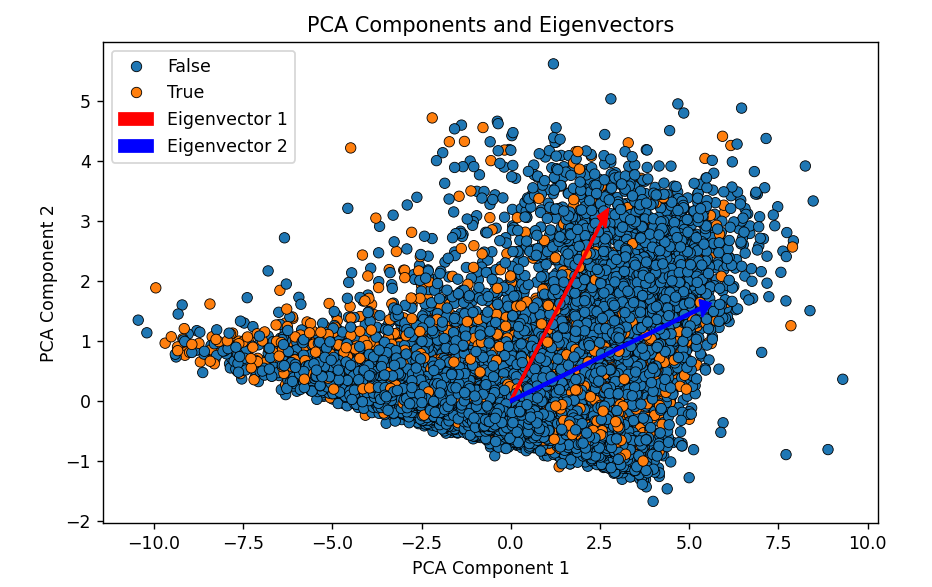}
        \caption*{(a)}
    \end{minipage}
    
    \vspace{0.5cm}
    
    \begin{minipage}{0.7\linewidth}
        \centering
        \includegraphics[width=\linewidth]{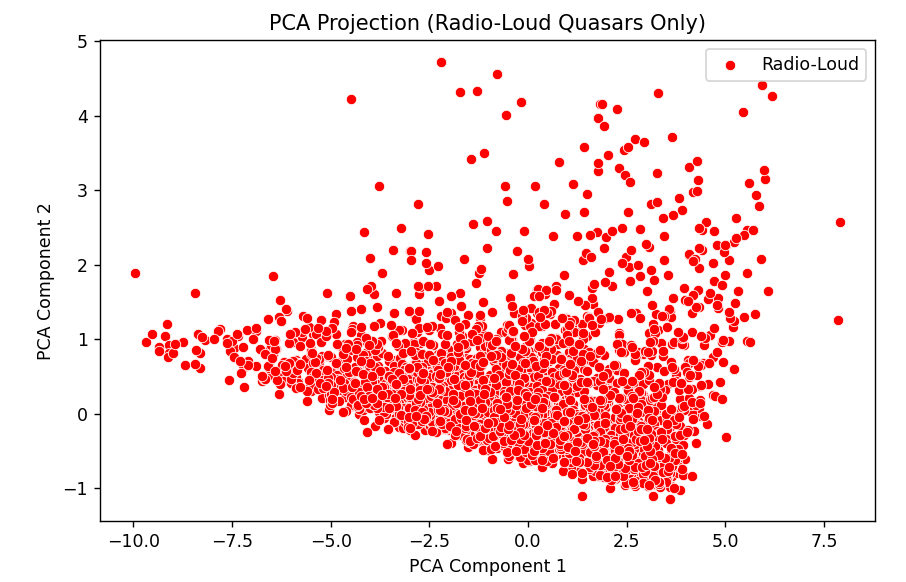}
        \caption*{(b)}
    \end{minipage}

    \begin{minipage}{0.7\linewidth}
        \centering
        \includegraphics[width=\linewidth]{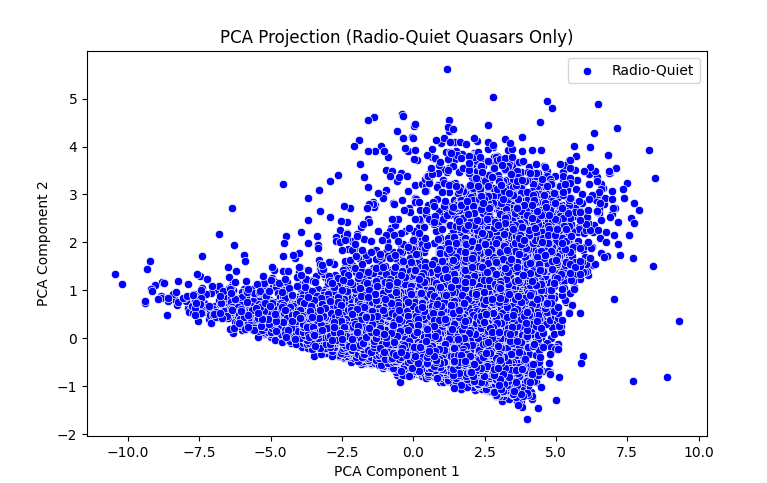}
        \caption*{(c)}
    \end{minipage}
    \caption{\small{Panel (a) shows dimensionally reduced data plotted against the first two PCA components. Panel (b) shows PCA projection of only radio-loud quasars. Panel (c) shows projection of radio-quiet quasars. }}
    \label{fig:pca_combined}
\end{figure}

Since the distribution of the RL and RQ quasars in the upper quadrant is significantly different, regression classifier can draw a decision boundary with corresponding probabilities assigned to the two regions separated by boundary thus helping in increasing the recall for RL class with balanced class weight. To evaluate the ability of different models to classify radio-loud and radio-quiet quasars, we compared fiur approaches: a Random Forest classifier trained on the original photometric features (\texttt{u\_mag}, \texttt{g\_mag}, \texttt{r\_mag}, \texttt{i\_mag}, \texttt{z\_mag}), a logistic regression classifier trained on the first two principal components derived from a PCA decomposition of the same features, an XGBoost classifier with threshold tuning trained on the original dataset and a Random Forest Classifier trained on a SMOTE transformed dataset. In this section, we discuss performance metrics for RFC and PBC as summarized in Table~\ref{tab:classifier-comparison}.\\

The Random Forest classifier achieved a high overall accuracy of 92\%, indicating strong predictive power for the majority class (radio-quiet quasars). However it is evident that the model is biased toward the majority class. The recall for radio-quiet quasars is 1.00, while the recall for radio-loud quasars is only 0.04, meaning that the model correctly identifies almost all radio-quiet quasars but fails to detect the minority class of radio-loud quasars. This is also reflected in the low precision and F1-score for the radio-loud class (0.44 and 0.08, respectively), highlighting poor sensitivity to the minority class despite high overall accuracy. The class imbalance is evident from the fact that the total number of radio quiet quasars are 42663 and radio loud quasars are only 3757 in number.

\begin{table}[H]
\centering
\caption{Comparison of Classification Metrics: Random Forest vs. PCA-based Logistic Regression}
\label{tab:classifier-comparison}
\begin{tabular}{lcccc}
\toprule
\textbf{Metric} & \textbf{Random Forest} & \textbf{PCA Logistic Regression (Balanced)} \\
\midrule
\textbf{Accuracy}             & 0.92  & 0.62 \\
\textbf{Precision (Radio-Quiet)} & 0.92  & 0.94 \\
\textbf{Precision (Radio-Loud)}  & 0.44  & 0.11 \\
\textbf{Recall (Radio-Quiet)}    & 1.00  & 0.63 \\
\textbf{Recall (Radio-Loud)}     & 0.04  & 0.52 \\
\textbf{F1-score (Radio-Quiet)}  & 0.96  & 0.76 \\
\textbf{F1-score (Radio-Loud)}   & 0.08  & 0.19 \\
\textbf{Macro Avg F1-score}      & 0.52  & 0.47 \\
\textbf{Weighted Avg F1-score}   & 0.88  & 0.71 \\
\bottomrule
\end{tabular}
\end{table}

On the other hand, the PCA-based logistic regression classifier (PBC), while achieving a lower overall accuracy of 62\%, shows more balanced performance across the two classes. The major challenge of class imbalance can be tackled in the PBC by using the balanced method from the logistic regression. Notably, the recall for radio-loud quasars increases significantly to 0.52, which implies that more than half of the actual radio-loud quasars are correctly identified. However, this improvement in sensitivity comes at the cost of reduced precision (0.11), suggesting a higher false positive rate for the radio-loud class. The radio-quiet recall also drops to 0.63, indicating reduced confidence in detecting the majority class. The macro-averaged F1-score of the PCA-based model is 0.47, slightly lower than the Random Forest's 0.52, reflecting a more even but less accurate performance overall. For this framework, if the primary objective is to reliably detect rare radio-loud quasars despite more false positives, the PCA-based model may be preferable. However, for applications where overall classification accuracy is more important and class imbalance is tolerable, Random Forest offers stronger performance. Future improvements may involve studying this critical tradeoff in detail to develop optimal classification strategies aiding in overall increased accuracy.\\

The ROC curve for both classifier methods performs slightly better than random classification assignment of radio-loud and radio-quiet quasars. In both classifiers, precision recall curves also drop quiet fast. However as expected, due to balanced regression, PBC maintains a gradual slope even at higher recall while the RFC curve drops faster at higher recalls. RFC achieves overall very less recall for the imbalanced class and is outperformed by PBC in that aspect.

\begin{figure}[H]
    \centering
    % Top row: RFC
    \begin{minipage}[b]{0.45\linewidth}
        \centering
        \includegraphics[width=\linewidth]{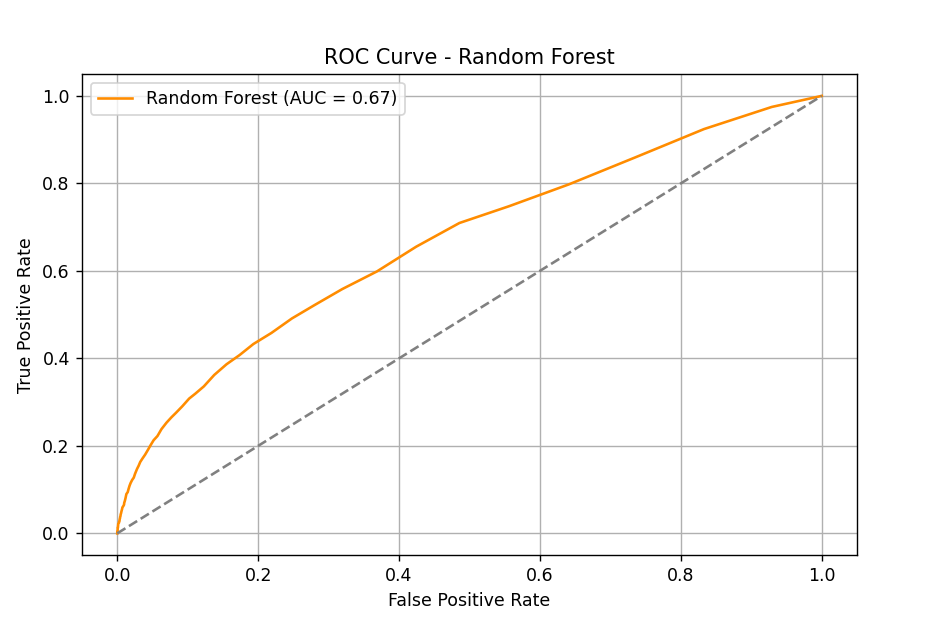}
        \textbf{(a)} ROC curve for RFC
    \end{minipage}
    \hfill
    \begin{minipage}[b]{0.45\linewidth}
        \centering
        \includegraphics[width=\linewidth]{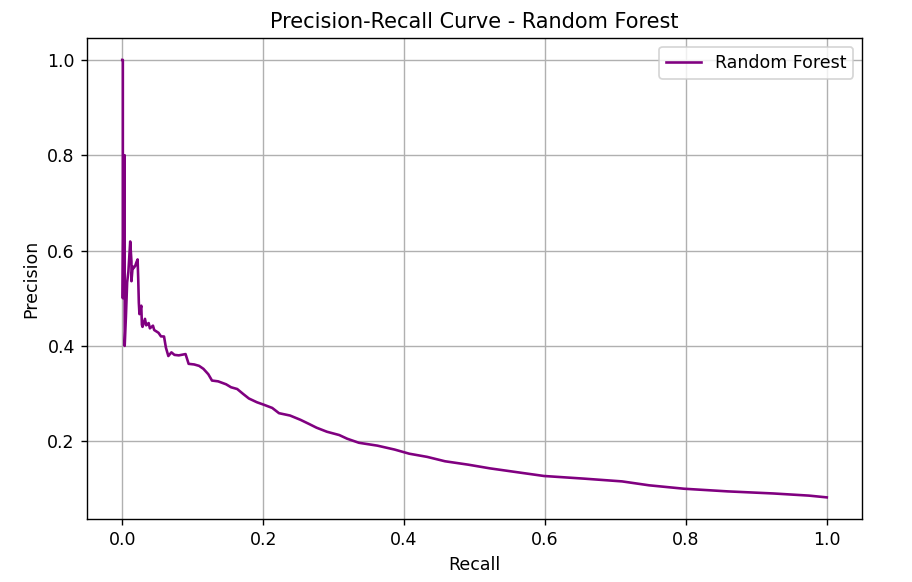}
        \textbf{(b)} Precision-Recall curve for RFC
    \end{minipage}

    \vspace{0.5cm}

    % Bottom row: PBC
    \begin{minipage}[b]{0.45\linewidth}
        \centering
        \includegraphics[width=\linewidth]{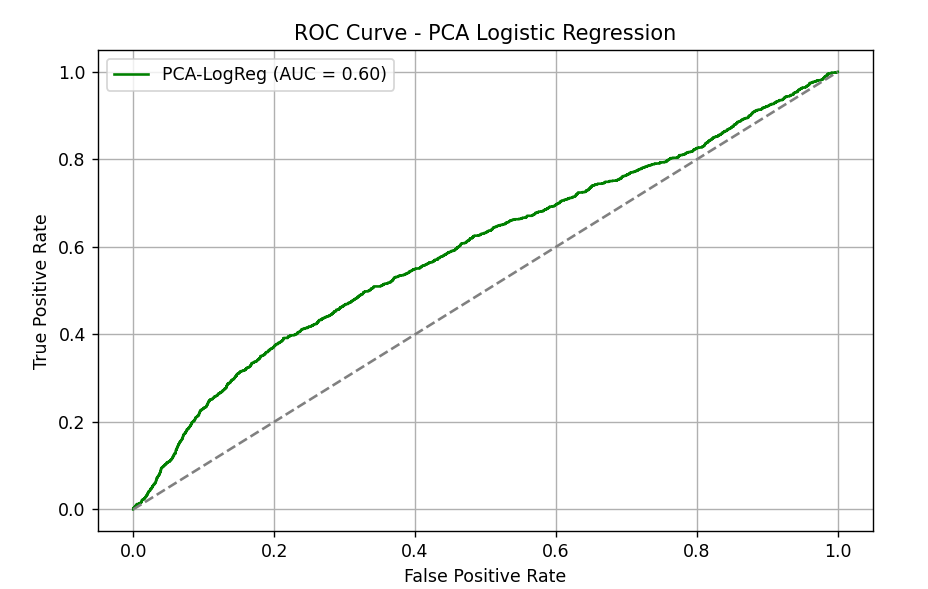}
        \textbf{(c)} ROC curve for PBC
    \end{minipage}
    \hfill
    \begin{minipage}[b]{0.45\linewidth}
        \centering
        \includegraphics[width=\linewidth]{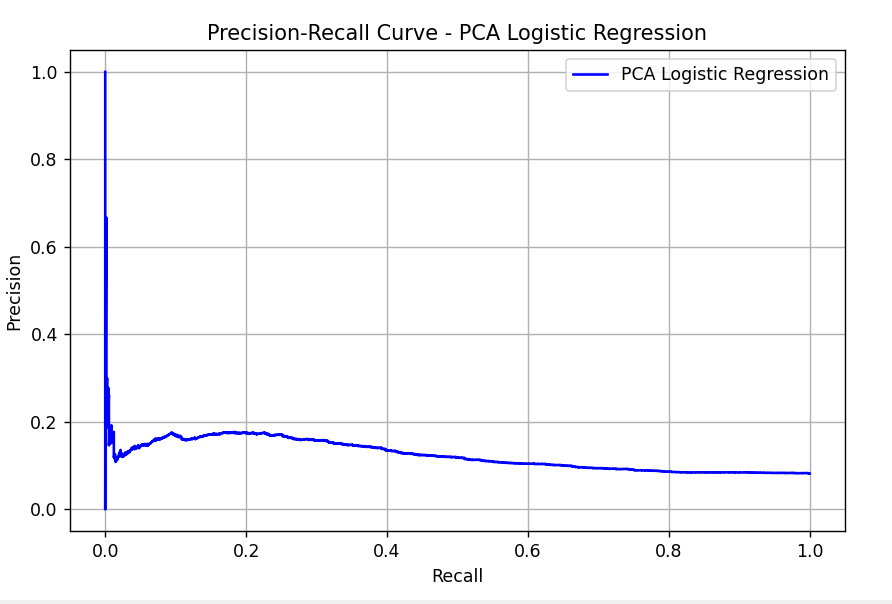}
        \textbf{(d)} Precision-Recall curve for PBC
    \end{minipage}

    \caption{Comparison of classification performance using ROC and Precision-Recall curves for RFC (top row) and PBC (bottom row).}
    \label{fig:comparison-rfc-pbc}
\end{figure}

 This is a classic behavior in case of class imbalance problems. As is established before, in general RFC outperforms PBC if class imbalance is not considered of central importance. Our model also performs better than RFC+SMOTE pipeline on the recall of the minority class objects.

\begin{table}[H]
\centering
\begin{tabular}{|l|c|c|}
\hline
\textbf{Metric} & \textbf{PBC} & \textbf{RFC + SMOTE} \\ \hline
Accuracy & 0.62 & 0.85 \\ \hline
Recall (Radio-Loud) & 0.52 & 0.25 \\ \hline
Precision (Radio-Loud) & 0.11 & 0.18 \\ \hline
F1 (Radio-Loud) & 0.19 & 0.21 \\ \hline
Macro Avg Recall & 0.58 & 0.57 \\ \hline
Weighted Avg F1 & 0.71 & 0.86 \\ \hline
\end{tabular}
\caption{Comparison of PBC and RFC + SMOTE Metrics}
\end{table}

 However both classifiers might perform better overall with application of SMOTE and other resampling techniques. Stacked models can also be studied with multiple techniques applied to find a sweet spot between the recall,precision,F1 score of minority class and overall accuracy of the prediction.

\subsection{XGBoost with threshold tuning model}

In a binary classification task using XGBoost, the model outputs a probability score $\hat{p} = P(y = 1 \mid \mathbf{x})$ for each input feature vector $\mathbf{x}$. To convert this probability into a predicted class label, a decision threshold $\theta$ is applied:

\[
\hat{y} =
\begin{cases}
1, & \text{if } \hat{p} \geq \theta \\
0, & \text{if } \hat{p} < \theta
\end{cases}
\]

By default, the threshold $\theta$ is set to 0.50. This means that:

\begin{itemize}
  \item If the predicted probability $\hat{p}$ is greater than or equal to 0.50, the model predicts the positive class (e.g., a radio-loud quasar).
  \item If the predicted probability $\hat{p}$ is less than 0.50, the model predicts the negative class (e.g., a radio-quiet quasar).
\end{itemize}

Adjusting the threshold $\theta$ allows one to balance between recall and precision, which is particularly important when dealing with imbalanced datasets. XGBoost trained data outperforms both the classifiers in terms of F1 score and precision at the threshold of 0.50. However, this comes at the cost of reduction in the recall of the minority class as compared to PBC. Overall accuracy of XGBoost is better than PBC but lesser than RFC. XGBoost provides a balanced approach between the overall accuracy, F1 score and precision of the minority class. 

\begin{table}[H]
\centering
\caption{XGBoost Classification Metrics at Various Thresholds}
\begin{tabular}{c|cccc|cccc|c}
\toprule
\textbf{Threshold} & \multicolumn{4}{c|}{\textbf{Radio-Quiet (False)}} & \multicolumn{4}{c|}{\textbf{Radio-Loud (True)}} & \textbf{Accuracy} \\
 & Precision & Recall & F1 & Support & Precision & Recall & F1 & Support & \\
\midrule
0.10 & 0.96 & 0.11 & 0.20 & 8537 & 0.09 & 0.95 & 0.16 & 747 & 0.18 \\
0.20 & 0.95 & 0.30 & 0.46 & 8537 & 0.09 & 0.82 & 0.17 & 747 & 0.35 \\
0.30 & 0.95 & 0.52 & 0.67 & 8537 & 0.11 & 0.67 & 0.19 & 747 & 0.53 \\
0.40 & 0.94 & 0.69 & 0.79 & 8537 & 0.12 & 0.51 & 0.20 & 747 & 0.67 \\
\textbf{0.50} & \textbf{0.94} & \textbf{0.81} & \textbf{0.87} & \textbf{8537} & \textbf{0.15} & \textbf{0.39} & \textbf{0.22} & \textbf{747} & \textbf{0.78} \\
0.60 & 0.93 & 0.90 & 0.91 & 8537 & 0.19 & 0.29 & 0.23 & 747 & 0.85 \\
0.70 & 0.93 & 0.95 & 0.94 & 8537 & 0.25 & 0.21 & 0.23 & 747 & 0.89 \\
0.80 & 0.93 & 0.98 & 0.95 & 8537 & 0.33 & 0.12 & 0.18 & 747 & 0.91 \\
0.90 & 0.92 & 1.00 & 0.96 & 8537 & 0.47 & 0.04 & 0.07 & 747 & 0.92 \\
\bottomrule
\end{tabular}
\end{table}

Threshold tuning gives an optimal performance matrix for XGBoost at the threshold of 0.50. This model can be preferred where a tradeoff between the recall of the minority class and precision is sensitive. To achieve a higher precision in prediction at the cost of predicting less minority class members, one can prefer XGBoost over other algorithms.

\begin{figure}[H]
    \centering
    \includegraphics[width=0.9\linewidth]{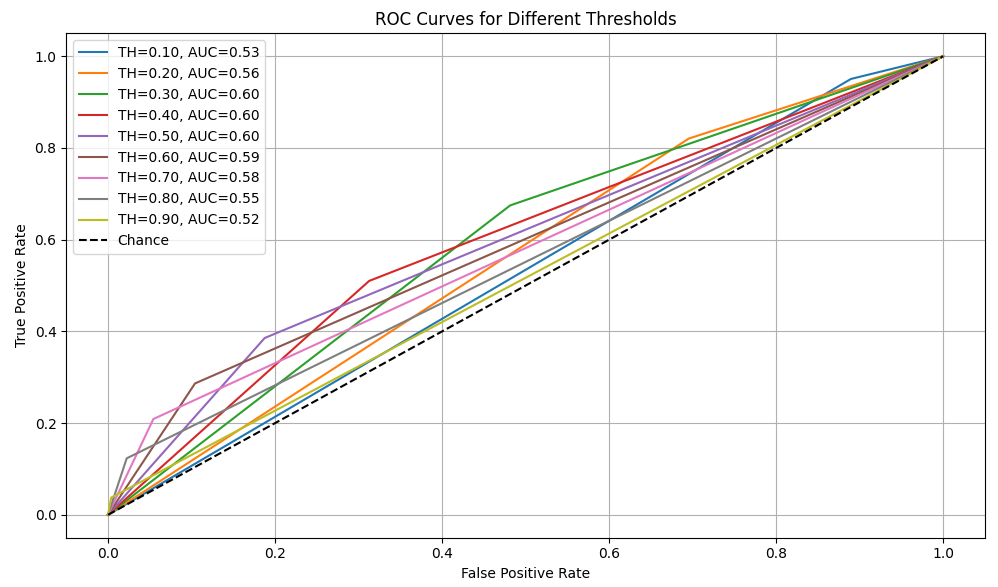}
    \caption{Figure shows ROC curves for different thresholds for the XGBoost treated dataset}
    \label{fig:enter-label}
\end{figure}

For threshold ROC curves best AUC of 0.60 is obtained at thresholds of 0.3,0.4 and 0.5. These AUCs are comparable to the PBC but lesser than RFC.

\begin{figure}[H]
    \centering
    \includegraphics[width=0.9\linewidth]{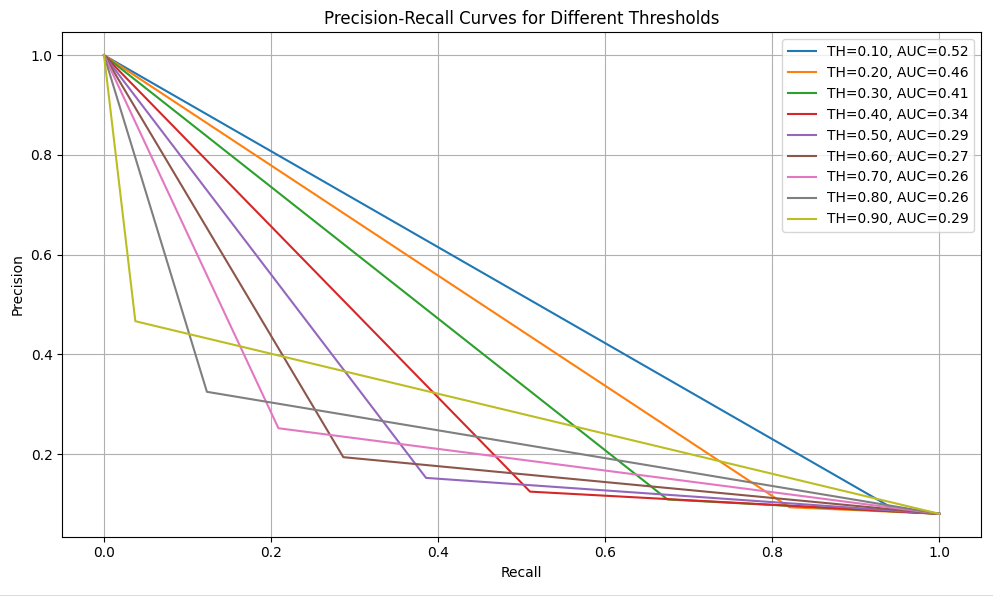}
    \caption{Precision recall curves for different threshold for the XGBoost treated dataset}
    \label{fig:enter-label}
\end{figure}

Precision recall curves highlight the central issue of rapidly falling precision, with increased recall for the minority class. Though a threshold of 0.10 performs ideally for the precision-recall, it severely undermines the other functionalities of the classifier. In the dataset, the ratio of the radio-loud to radio-quiet quasars is nearly 1:14 which shows the severe class imbalance. This class imbalance is handled in our model by setting the scale factor at the class imbalance ratio which inherently takes into account the extent of the class imbalance. However the recall for the minority class at the optimal threshold is still bad. This recall can be increased by adjusting the scale factor more aggressively so that the model gives more attention to the minority class. This will come at the cost of reduced overall prediction accuracy and one may have to use additional SMOTE and other resampling techniques to understand better the more optimal solution to the classification problem.

\section*{Discussion}

In this work, we have systematically analyzed the high-dimensional Sloan Digital Sky Survey (SDSS) quasar dataset to distinguish between radio-loud and radio-quiet quasars, as well as to identify high-redshift quasars. We implemented and compared the performance of three different classification models: a Random Forest Classifier (RFC), XGBoost model with threshold tuning and a PCA-based balanced logistic regression classifier (PBC).\\

By applying Principal Component Analysis (PCA) to the feature space, we demonstrate that the first two principal components are sufficient to capture the majority of the variance in the dataset. This dimensionality reduction improves computational efficiency. Our analysis highlights the ability of PCA to select dominant features contributing to class separation. One caveat of this methodology is that PCA trained data may not clearly maintain the original structure of the dataset. PCA focuses on retaining the most variance but not necessarily the characteristic functionalities of data variables. The new uncorrelated PCA components may provide very less information making XGBoost, SMOTE or other techniques futile. Any stacked model prepared is likely to perform worse with PCA applied before training the model. Hence, though in case of regression model the PCA works, it may not work effectively for other stacked models. This once again highlights the seriousness of choosing an accurate model based on the goal of the analysis.\\

In this analysis, RFC achieves a highest overall accuracy in classification; but with a severe limitation in identifying radio-loud quasars which are a minority class. In contrast, the PCA-based logistic regression classifier, though slightly lower in overall accuracy, provides a significantly improved recall for the minority class. XGBoost on the other hand, handles the F1 score and the precision very well but at the cost of lesser overall accuracy than RFC and lesser recall for minority class than PBC.

\begin{table}[H]
\centering
\caption{Performance comparison of XGBoost (Threshold = 0.50), Random Forest Classifier (RFC), PCA-based Logistic Regression (PBC), and RFC + SMOTE. Bolded and underlined values indicate the best performance per metric.}
\label{tab:model-comparison}
\begin{tabular}{lcccc}
\toprule
\textbf{Metric} & \textbf{XGBoost (TH=0.50)} & \textbf{RFC} & \textbf{PBC} & \textbf{RFC + SMOTE} \\
\midrule
Accuracy & 0.78 & \underline{\textbf{0.92}} & 0.62 & 0.85 \\
Precision (Radio-Quiet) & \underline{\textbf{0.94}} & \underline{\textbf{0.94}} & \underline{\textbf{0.94}} & 0.93 \\
Precision (Radio-Loud) & 0.15 & \underline{\textbf{0.44}} & 0.11 & 0.18 \\
Recall (Radio-Quiet) & 0.81 & \underline{\textbf{1.00}} & 0.63 & 0.90 \\
Recall (Radio-Loud) & 0.39 & 0.04 & \underline{\textbf{0.52}} & 0.25 \\
F1-score (Radio-Quiet) & 0.87 & \underline{\textbf{0.96}} & 0.76 & 0.92 \\
F1-score (Radio-Loud) & \underline{\textbf{0.22}} & 0.08 & 0.19 & 0.21 \\
Macro Avg F1-score & {{0.54}} & 0.52 & 0.47 & \underline{\textbf{0.56}} \\
Weighted Avg F1-score & 0.82 & \underline{\textbf{0.88}} & 0.71 & 0.86 \\
\bottomrule
\end{tabular}
\end{table}

The performance comparison matrix demonstrates that each of the classifier methods has its own tradeoffs to improve a certain aspect of classification. This trade-off between overall accuracy, F1 score and class-specific recall emphasizes the importance of model selection based on the analysis objective. The novel methodological pipeline of PCA followed by a balanced regression classifier provides valuable insights into increasing the recall for the minority class while balancing the overall accuracy. Our model also maintains a F1 score higher than RFC and comparable to XGBoost and SMOTE+RFC pipeline thus indicating its potential in addressing class imbalance problems. Our model provides the best recall for minority class in all of the other models.\\

Additionally, we explore the Lyman-alpha Forest absorption effect, particularly visible in the $u$ and $g$ band magnitudes for high-redshift quasars (typically $z > 3$). This effect is an important signature in identification of such QSOs. We further support our findings with Kernel Density Estimation (KDE) techniques to visualize the redshift distribution under two distinct likelihood models, offering insight into the underlying population structure of quasars.\\

Overall, our integrated framework demonstrates the effectiveness of combining dimensionality reduction, classification, and statistical analysis techniques for astrophysical data mining. This approach enhances the classification task on SDSS quasar data and can be applied to present and future high dimensional, high volume surveys. Future work would focus on application of SMOTE and other minority class resampling techniques to improve the performance of the tested classifiers and introduce and test new SVM and neural network based classifiers for enhanced performance.

\end{document}